\documentclass[10pt]{article}
\usepackage{fullpage,graphicx,subfigure,mathdots,mathpazo,color}
\usepackage{amsmath,amscd,tikz,mathrsfs,cite}
\usepackage[normalem]{ulem}
\usepackage{amsmath}
\usepackage{setspace}

\usepackage{epsfig,amsmath,graphicx,amssymb,overpic}
\usepackage{bm}
\usepackage{graphicx}
\usepackage{subfigure, xcolor}
\usepackage{ntheorem}
\usepackage{listings,diagbox}

\usepackage{color}
\usepackage{epsfig,amsmath,graphicx,amssymb,overpic}

\usepackage{booktabs}
\usepackage{diagbox}
\usepackage{graphicx}
\usepackage{dcolumn}
\usepackage{bm}
\usepackage{graphicx}
\usepackage{subfigure}
\usepackage{epsfig,amsmath,graphicx,amssymb,overpic,cite}
\usepackage{makecell}

\def\be{\begin{equation}}
\def\ee{\end{equation}}
\def\bee{\begin{eqnarray}}
\def\ene{\end{eqnarray}}
\def\bes{\begin{subequations}}
\def\ees{\end{subequations}}

\newcommand{\PT}{{\cal PT}}

\def\v{\vspace{0.1in}}

\allowdisplaybreaks[4]

\begin{document}

\baselineskip=14pt \renewcommand {\thefootnote}{\dag}
\renewcommand
{\thefootnote}{\ddag} \renewcommand {\thefootnote}{ }

\pagestyle{plain}

\begin{center}
\baselineskip=16pt \leftline{} \vspace{-.3in} {\Large \textbf{Symmetry breaking bifurcations and excitations of solitons
in linearly coupled NLS equations with $\mathcal{PT}$-symmetric potentials}} \\[0.2in]

Jin Song$^{1,2}$, Boris A. Malomed$^{3,4}$, and Zhenya Yan$^{{1,2},{*}}$%
\footnote{$^{*}$Corresponding author. \textit{Email address}:
zyyan@mmrc.iss.ac.cn} \\[0.15in]
\textit{{\small $^{1}$KLMM, Academy of Mathematics and Systems Science,
Chinese Academy of Sciences, Beijing 100190, China \\[0pt]
$^{2}$School of Mathematical Sciences, University of Chinese Academy of
Sciences, Beijing 100049, China\\[0pt]
$^{3}$Department of Physical Electronics, School of Electrical Engineering,
Faculty of Engineering, Tel Aviv University, \\ Tel Aviv 69978, Israel \\[0pt]
$^{4}$Instituto de Alta Investigaci\'{o}n, Universidad de Tarapac\'{a},
Casilla 7D, Arica, Chile }} \\[0pt]
\end{center}

\noindent \textbf{Abstract:}\thinspace\ We address symmetry breaking bifurcations
(SBBs) in the ground-state (GS) and dipole-mode (DM) solitons of the 
1D linearly coupled NLS equations, modeling the propagation of light in a dual-core planar waveguide with the
Kerr nonlinearity and two types of $\mathcal{PT}$-symmetric
potentials. The $\PT$-symmetric potential is employed to obtained different types of solutions.
A supercritical pitchfork bifurcation
occurs in families of symmetric solutions of both the GS and DM types. A novel feature of the system is interplay
between breakings of the $\mathcal{PT}$ and inter-core symmetries. Stability
of symmetric GS and DM modes and their asymmetric counterparts, produced by
SBBs of both types, is explored via the linear-stability analysis and simulations. It is found that the instability of $\mathcal{PT}$%
-symmetric solutions takes place prior to the inter-core symmetry breaking.
Surprisingly, stable inter-core-symmetric GS solutions may remain stable
while the $\mathcal{PT}$ symmetry is broken. Fully asymmetric GS and DM
solitons are only partially stable. Moreover, we construct symmetric and asymmetric GS
solitons under the action of a pure imaginary localized potential, for which
the SBB is subcritical.
These results exhibit that stable solitons can still be found in dissipative systems.
Finally, excitations of symmetric and
asymmetric GS solitons are investigated by making the potential's parameters or
the system's coupling constant functions,
showing that GS solitons can be converted from an asymmetric shape onto a
symmetric one under certain conditions.
These results may pave the way for the study of linear and nonlinear phenomena in a dual-core planar waveguide with $\mathcal{PT}$
potential and related experimental designs.

\section{Introduction}

A commonly known principle of classical and quantum physics is that the
ground state (GS) of linear systems exactly follows the symmetry of the
underlying Hamiltonian, while excited states may realize other
representations of the same symmetry. In particular, the quantum-mechanical
GS in a symmetric potential well is always spatially even, while the first
excited state is represented an odd wave function \cite{LL}. The same is
true for the linear propagation of light in waveguides, in which an
effective potential is induced by a spatial profile of the local refractive
index, as the paraxial transmission of light in the waveguide is governed by
the linear equation which has the same form as the linear Schr\"{o}dinger
equation in quantum mechanics \cite{Agrawal}.

In nonlinear optical waveguides made of a self-focusing material, the wave
propagation obeys the symmetry of the effective guiding potential only in
the regime of weak nonlinearity, which is quantified by small values of the
power of the optical beam. A generic effect which follows the increase of
the power is the symmetry-breaking bifurcation (SBB), which makes the shape
of the respective GS wave mode spatially asymmetric \cite{book}. In
particular, because the SBB gives rise to a degenerate pair of two GSs,
which are mirror copies of each other, the transition to the asymmetric GS
implies that the fundamental principle of quantum mechanics, which does not
allow degeneracy of the GS \cite{LL}, is no longer valid for nonlinear
modes. Above the SBB point, a spatially symmetric mode coexists with the
asymmetric GSs, but it is unstable against symmetry-breaking perturbations.

An essential ramification of the phenomenology of the spontaneous symmetry
breaking is provided by SBB in dual-core waveguides, such as twin-core
optical fibers, with the cubic self-focusing acting in each core~\cite{Jensen,Viskol}:
\begin{equation}
\begin{split}
i\phi _{z}& =-\frac{1}{2}\phi_{xx}-|\phi |^{2}\phi + U(x)\phi -\kappa \psi, \\
i\psi _{z}& =-\frac{1}{2}\psi_{xx}-|\psi |^{2}\psi +U(x)\psi -\kappa \phi,
\end{split}
\label{phipsi-0}
\end{equation}
where $\phi \equiv \phi (x,z)$ and $\psi \equiv \psi (x,z)$ are slowly
varying amplitudes of the electromagnetic waves with the transverse coordinate $x$ and propagation distance $z$,
$U(x)$ represents the real-valued external potential well (e.g., $U(x)=x^2/2$), $\kappa$ is  the linear-coupling
constant. In these systems, the interplay between the intra-core
self-focusing ($-|\phi |^{2}\phi$ or $-|\psi |^{2}\psi$) and
group-velocity dispersion ($-\frac{1}{2}\phi_{xx}$ or $-\frac{1}{2}\psi_{xx}$),
and the inter-core linear coupling ($-\kappa \psi$ or $-\kappa \phi$), provided by tunneling of light across the barrier separating the
guiding core, gives rise to the inter-core SBB. The bifurcation replaces the
original GS, which was symmetric with respect to the parallel cores, by a
spontaneously established asymmetric one with unequal powers carried by the
cores. The SBB of this type was studied in detail theoretically  \cite{Wabnitz,Pare,Maim,Snyder,Akhmed,Pak} and recently demonstrated experimentally in a twin-core
optical fiber (coupler) \cite{Bugar}.
For example, symmetric and asymmetric soliton states in twin-core nonlinear optical fibers were
examined using an improved variational approximation~\cite{Pak}.
Depending on the wave form, which may
be delocalized or self-trapped, the SBB in the dual-core system may be of
the supercritical (forward) or subcritical (backward) type, according to the
usual definition \cite{bif}. In either case, the bifurcation give rise to
the destabilization of the state which is symmetric with respect to the
parallel cores and creation of a pair of asymmetric ones. The SBBs of the
sub- and supercritical types give rise, respectively, to branches of the
asymmetric solutions which evolve backward or forward, as stable or unstable
ones, with respect to the variation of the full power of the optical beam.
In the subcritical case, the unstable branches reach turning points and
revert to the forward direction, getting stable as a result of the
reversion. Thus, the respective stable asymmetric states emerge
subcritically, at a value of the power which corresponds to the turning
points, being smaller than the power corresponding to the SBB point.
Accordingly, the full set of stationary states is bistable in the interval
between the turning and SBB points, where the stable asymmetric states
coexist with the symmetric ones, which still remain stable in this interval.
Furthermore, the relationships between dispersion effects and soliton dynamics were also
discussed \cite{hou1,hou2,hou3,hou4} such that some soliton solutions of  coupled NLS equations~\cite{hou1},
and the modulation instability was found  in nonlinear optical fibers  higher-order dispersions~\cite{hou4}.

Symmetry-breaking effects were also studied in dual-core laser fibers, which
include gain and loss in addition to the intra-core group-velocity
dispersion and self-focusing nonlinearity, and inter-core linear coupling.
The model of the dual-core fiber is based on a pair of linearly coupled
complex Ginzburg-Landau equations \cite{Sigler}. A spontaneously established
asymmetric regime of the operation of coupled lasers was observed in the
experiment \cite{lasers}. The model of the dual-core fiber including the gain and loss suggests one a
possibility to consider a symmetric system with a parity-time ($\mathcal{PT}$%
)-symmetric distribution of spatially separated local gain and loss in each
core. As is well known, linear spectra of the $\mathcal{PT}$-symmetric
systems remain purely real due to the exact balance between the gain and
loss, provided that the strength of the gain-loss distribution does not
exceed a certain critical value \cite{pt11,pt22,pt4,pt7,Bender-review,Chr-review,review1,review2}.
And a multitude of intensive research results on $\PT$-symmetric systems have
been reported \cite{re5,re6,re7,re8,re9,re9b,re9c,re10,re11,re12,re13,re14,SSB2,SSB7,SSB8,daili}.
Experimentally, setups demonstrating persistent $\mathcal{PT}$ symmetry were
realized in optics \cite{pt7,Segev,exp1}.

Previously, $\mathcal{PT}$-symmetric dual-core systems were considered with
the gain and loss of equal strengths carried by the different cores \cite%
{PT1,PT2}. In that system, the breakup of the $\mathcal{PT}$ symmetry, i.e.,
a transition from a purely real spectrum to one including complex
eigenvalues, takes place when the strength of the gain and loss becomes
equal to the constant of the inter-core coupling.
In fact, the separation of
the gain and loss between the two cores breaks the exact symmetry between
them, even if the $\mathcal{PT}$ symmetry stays unbroken.

In the present work, we aim to elaborate a $\mathcal{PT}$-symmetric
dual-core system of a different type, which keeps the full symmetry between
the cores, each carrying the self-focusing cubic nonlinearity and a $%
\mathcal{PT}$-symmetric potential, i.e., a complex one whose real and
imaginary parts are spatially even and odd, respectively. 
The main contributions of this paper can be concluded as follows:
\begin{itemize}

\item{} We introduce a model, based on the linearly-coupled NLS equations,
each including the complex potential, and consider the potentials with two different imaginary parts, \textit{viz}.,
spatially localized or delocalized ones. To the best of our knowledge, models of this type were not studied previously.

\item{} We discuss the interplay between the nonlinearity-induced spontaneous breaking of the inter-core
symmetry and breakup of the $\mathcal{PT}$ symmetry in individual cores, and numerically find that SBBs of the pitchfork type account for the breaking of the inter-core symmetry in the spatially even GS
solutions and in odd ones which represent dipole modes (DMs, i.e., the first
excited states admitted by the intra-core potentials).

\item{} Unlike the even GS soliton, dipole mode does not exist in the absence of the harmonic-oscillator potential.
And as the strength of gain and loss changes, the solitons exhibit different states.
The breakup of the $\mathcal{PT}$ symmetry occurs before the breaking of the inter-core
symmetry.
A noteworthy finding is that stable inter-core-symmetric GS
solutions keep their stability in a small region even when the $\mathcal{PT}$
symmetry is broken.

\item{} An essential difference from the well-known properties
of the nonlinear coupler without the gain and loss terms is that inter-core
asymmetric GS and DM modes are only partly stable, as some of them are
destabilizes by the $\mathcal{PT}$-symmetry breaking.

\item{} For the system with purely imaginary potentials,
the SBB changes its character from
supercritical to subcritical.

\item{} We investigate the transforms between symmetric and asymmetric GS modes in a nonstationary
system, in which parameters of the $\mathcal{PT}$-symmetric potential and/or
the inter-core coupling constant are defined as functions of the propagation
distance. In particular, it is found that asymmetric GS modes can be
transformed back into symmetric ones by means of a properly defined modulation.
\end{itemize}

The rest of this paper is arranged as follows. Sec. 2 introduces a model, based
on the linearly-coupled NLS equations,
each including the complex potential with two different imaginary parts, \textit{viz}.,
spatially localized or delocalized ones. In Sec. 3, we mainly discuss the interplay
between the nonlinearity-induced spontaneous breaking of the inter-core
symmetry and breakup of the $\mathcal{PT}$ symmetry in individual cores.
In Sec. 4, we address the SSB in the system with purely imaginary potentials (without the real part).
In Sec. 5, transformations between symmetric and asymmetric GS modes are investigated in a nonstationary (modulated)
system. In Sec. 6,  we give the physical interpretations for all figures.
And the paper is concluded and discussed by Sec. 7.

\section{The dual-core system with $\mathcal{PT}$-symmetric potentials}

\subsection{The physical model and stationary soliton solutions}

The 1D twin-core optical fibers, with the cubic self-focusing acting in each core have been studied~\cite{Jensen,Viskol}.
Here we consider the 1D system of linearly coupled NLS equations, modeling the propagation of
light in a dual-core planar waveguide with the transverse coordinate $x$,
propagation distance $z$, intra-core Kerr nonlinearity, and the $\mathcal{PT}
$-symmetric potential is written as
\begin{equation}
\begin{split}
i\phi _{z}& =-\frac{1}{2}\phi _{xx}-|\phi |^{2}\phi +\left[ V(x)+iW(x)\right]
\phi -\lambda \psi , \\
i\psi _{z}& =-\frac{1}{2}\psi _{xx}-|\psi |^{2}\psi +\left[ V(x)+iW(x)\right]
\psi -\lambda \phi ,
\end{split}
\label{phipsi}
\end{equation}%
where $\phi \equiv \phi (x,z)$ and $\psi \equiv \psi (x,z)$ are slowly
varying complex amplitudes of the electromagnetic waves. The complex-valued
potential $V(x)+iW(x)$ is the $\mathcal{PT}$-symmetric one provided that its
real and imaginary parts are, respectively, even and odd functions, i.e., $%
V(-x)=V(x)$ and $W(-x)=-W(x)$. In fact, the $\PT$-symmetric external potential can be realized via
a judicious inclusion of loss or gain domains in guided wave geometries~\cite{pt7,pt}.
 We choose the real part as the
harmonic-oscillator potential, $V(x)=\frac{1}{2}x^{2}$, which is relevant as
the trap in many physical realizations, while the imaginary part with
strength $W_{0}$ and intrinsic scale $l$ can be naturally chosen as the spatially
delocalized one
\begin{equation}
W_{1}(x)=W_{0}\tanh \left( \frac{x}{l}\right) ,  \label{W1}
\end{equation}%
or spatially localized one
\begin{equation}
W_{2}(x)=W_{0}\,\mathrm{sech}\left( \frac{x}{l}\right) \tanh \left( \frac{x}{%
l}\right).  \label{W2}
\end{equation}%
%
%
%
%
%
%
%
%
Real parameter $\lambda >0$ in Eq. (\ref{phipsi}) is the inter-core coupling
coefficient which couples the parallel cores. Further, we assume that the
strengths of the self-focusing nonlinear terms are equal in both cores,
being fixed equal to $1$ by means of rescaling. We stress that, while
imaginary potential (\ref{W1}) does not vanish ($W_1(x)\rightarrow \pm W_0$) at $|x|\rightarrow \infty $,
it may have a chance to support stable modes due to the presence of the
strong confining potential $\sim x^{2}$, cf. Ref. \cite{Thawatchai}.

With propagation distance $z$ replaced by time $t$, Eq.~(\ref{phipsi}) plays
the role of the generalized Gross-Pitaevskii (GP) equation for Bose-Einstein
condensates (BECs) loaded in the $\mathcal{PT}$-symmetric potential \cite%
{QM,review2}. Furthermore, Eq.~(\ref{phipsi}) can be rewritten in the
variational form $i\{\phi ,\psi \}_{z}=\delta H/(\delta \{\phi ^{\ast },\psi
^{\ast }\})$ with the generalized complex-values Hamiltonian:
\begin{equation}
\displaystyle H=\int_{-\infty }^{+\infty }\Big[\frac{1}{2}(|\phi
_{x}|^{2}+|\psi _{x}|^{2})-\frac{1}{2}(|\phi |^{4}+|\psi
|^{4})+(V(x)+iW(x))(|\phi |^{2}+|\psi |^{2})-\lambda (\phi ^{\ast }\psi
+\phi \psi ^{\ast })\Big]\mathrm{d}x,
\label{Hal}
\end{equation}%
where $\ast $ represents the complex conjugate.

Solutions produced by Eq. (\ref{phipsi}) are characterized by their total
optical power (in the application to BECs, it is the total norm),
\begin{equation}
P(z)=P_{1}(z)+P_{2}(z)\equiv \int_{-\infty }^{+\infty }|\phi
(x,z)|^{2}dx+\int_{-\infty }^{+\infty }|\psi (x,z)|^{2}dx.  \label{P}
\end{equation}%
It follows from Eq.~(\ref{phipsi}) that the evolution equation for the power
is
\begin{equation}
\frac{dP(z)}{dz}=-2\int_{\mathbb{R}}W(x)(|\phi |^{2}+|\psi
|^{2})dx.  \label{dp}
\end{equation}%
Note that the power is conserved for solutions in which local powers $|\phi
(x)|^{2}$ and $|\psi (x)|^{2}$ are even functions of $x$, as the imaginary
part $W(x)$ of the complex potential is an odd function, due to the
condition of the $\mathcal{PT}$ symmetry.

Stationary solutions of Eq.~(\ref{phipsi}) are looked for in the usual form,
\begin{equation}
\left\{ \phi (x,z),\psi (x,z)\right\} =\left\{ u(x),v(x)\right\}e^{-i\mu z},
\label{stationary}
\end{equation}%
where real $-\mu $ stands for the propagation constant ($\mu $ is the
chemical potential in the GP version of the model), and stationary wave
functions $\left\{ u(x),v(x)\right\}$ vanish at $x\rightarrow \pm \infty $. Substituting ansatz (\ref%
{stationary}) in Eq.~(\ref{phipsi}) yields the following coupled ordinary
differential equations:
\begin{equation}
\begin{split}
\mu u=& -\frac{1}{2}\frac{d^{2}u}{dx^{2}}-|u|^{2}u+(V(x)+iW(x))u-\lambda v,
\\
\mu v=& -\frac{1}{2}\frac{d^{2}v}{dx^{2}}-|v|^{2}v+(V(x)+iW(x))v-\lambda u.
\end{split}
\label{uv}
\end{equation}%
Symmetric and antisymmetric solutions, with $u=v$ or $u=-v$, respectively,
obey the single equation
\begin{equation}
\left( \mu \pm \lambda \right) u=-\frac{1}{2}\frac{d^{2}u}{dx^{2}}%
-|u|^{2}u+(V(x)+iW(x))u.  \label{single}
\end{equation}

The Lagrangian corresponding to Eq.~(\ref{uv}) is shown as follows
\begin{equation}\label{L}
  \begin{split}
   \displaystyle L =&\int_{\mathbb{R}}\left[\frac{1}{2}(|u_x|^2+|v_x|^2)-\frac{1}{2}(|u|^4+|v|^4)+(V(x)+iW(x))(|u|^2+|v|^2)-\lambda(u^*v+uv^*)\right.\\
      &\left. \qquad -\mu(|u|^2+|v|^2) \right]\mathrm{d}x.
  \end{split}
\end{equation}
Notice that the stationary solutions (\ref{stationary}) can be studied analytically in terms of the variational approximation related to Eq.~(\ref{L})~\cite{VA}, and numerically via the modified squared-operator iteration method~\cite{yjk,msom}.
Furthermore, if one desires to investigate the temporal progression of approximate wave functions characterized by generalized time-dependent collective coordinates in the presence of a complex potential, then the dissipation functional formalism
was commonly employed \cite{va1}. In a specific case involving a $\PT$-symmetric potential, the evolution of the collective coordinates was examined, leading to the identification of an instability criterion for a ground state solution. Subsequently, this formalism, along with the aforementioned stability criterion, was utilized to analyze the evolution of solitons in the coupled NLS equations subject to potentials exhibiting both even and odd-$\PT$ characteristics \cite{va2,va3}.

Stability of stationary solutions (\ref{stationary}) is a crucially
important issue. In particular, antisymmetric states, with $u=-v$, are
definitely unstable, as $\lambda >0$ implies that they have a positive,
rather than negative, coupling energy.

The bifurcations breaking the symmetry between the components of the
solution in the two cores takes place when total power $P$ of the symmetric
state exceeds a critical value, i.e., at $P>P_{\mathrm{cr}}$. The asymmetry
of states with unequal powers in the two cores, $P_{1}\neq P_{2}$, is
characterized by parameter
\begin{equation}
\theta =\frac{|P_{1}-P_{2}|}{P_{1}+P_{2}}.  \label{asym}
\end{equation}%
On the other hand, the breaking of the $\mathcal{PT}$ symmetry occurs when
strength $W_{0}$ of gain-loss distribution, represented by the imaginary
potential $W(x)$, attains a certain threshold value.
A problem relevant to
the system under the consideration, which was not addressed in previous
works, is to find $P_{\mathrm{cr}}$ as functions of $W_{0}$.

\begin{figure}[!t]
\centering
{\scalebox{0.78}[0.78]{\includegraphics{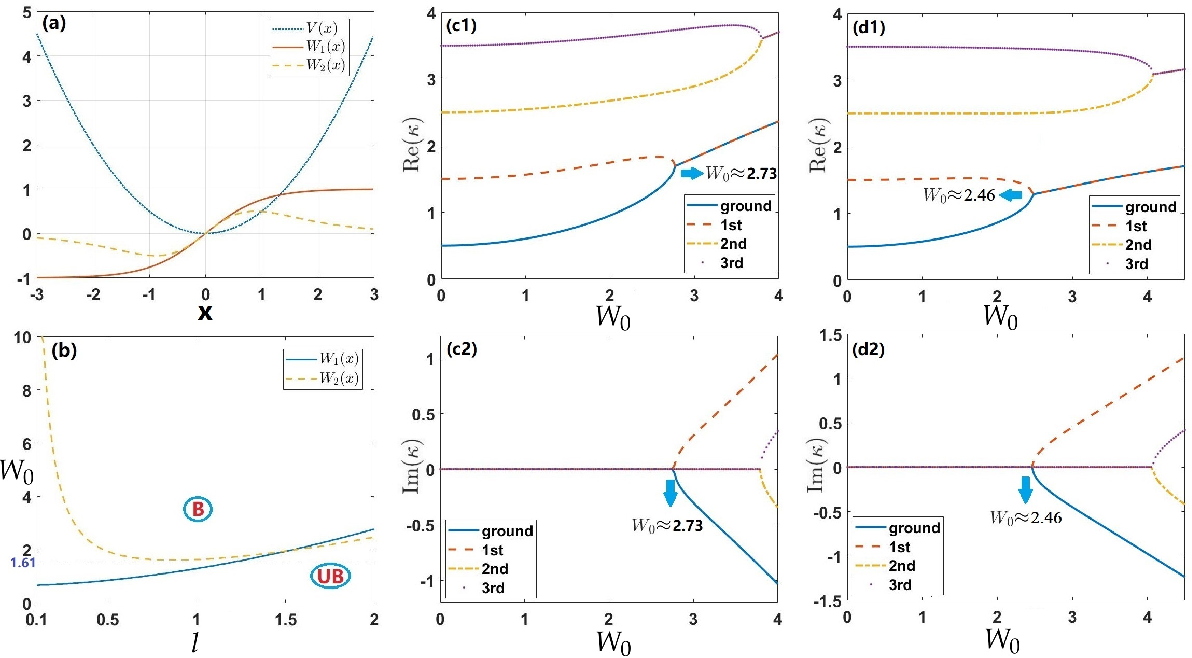}}} \hspace{-0.4in} \vspace{0.1in}
\caption{{\protect\small $\mathcal{PT}$ symmetry breaking for the linear spectral problem (\ref{linears}) with differential
gain-and-loss distributions given by Eqs.~(\ref{W1}) and (\ref{W2}).
 (a) Profiles of real and imaginary parts of the $%
\mathcal{PT}$-symmetric potentials (\protect\ref{W1}) and (\protect\ref{W2})
at $W_{0}=l=1$. (b) Results of the solution of the linear spectral problem (%
\protect\ref{linears}) for potentials $W_{1}(x)$ and $W_{2}(x)$, where
``UB" (``B") denotes the area of the
unbroken (broken) $\mathcal{PT}$ symmetry below (above) the breaking lines,
respectively. (c1,c2) Real and imaginary parts of the first two eigenvalues $%
\protect\kappa $ as functions of $W_{0}$ at $l=2$ for $W_{1}(x)$, where the $%
\mathcal{PT}$ symmetry breaking occurs at $W_{0}\approx 2.73$. (d1,d2) Real
and imaginary parts of the first two eigenvalues $\protect\kappa $ as
functions of $W_{0}$ at $l=2$ for $W_{2}(x)$, where the $\mathcal{PT}$
symmetry breaking occurs at $W_{0}\approx 2.46$.}}
\label{pt}
\end{figure}

\subsection{The linear spectral problem and $\mathcal{PT}$-symmetry breaking}

The calculation of the linear spectrum of the single equation (\ref{single})
with $\lambda =0$ is a well-known basic problem in the theory of linear $%
\mathcal{PT}$-symmetric systems. Here, we first consider the $\mathcal{PT}$%
-symmetry breaking for the following linear non-Hermitian Hamiltonian
related to system (\ref{uv}):
\begin{equation}\label{linears}
  \left(
       \begin{array}{cc}
       \mathcal{H} & -\lambda \\
       -\lambda    & \mathcal{H}
       \end{array}
  \right)\left(
              \begin{array}{c}
              \hat{u} \\
              \hat{v}
              \end{array}
          \right)=\kappa\left(
              \begin{array}{c}
              \hat{u} \\
              \hat{v}
              \end{array}
          \right),\quad \mathcal{H}=-\frac{1}{2}\partial_x^2+\frac{1}{2}x^2+iW(x),
\end{equation}
where $\kappa $ and $(\hat{u},\hat{v})^{T}$ are the eigenvalue and localized
eigenfunction, respectively. Although $\lambda $ affects the spectrum of the
linear eigenvalue problem (\ref{linears}), the $\mathcal{PT}$ symmetry
breaking is independent of $\lambda $, i.e., $\lambda $ has no effect on the
occurrence of complex eigenvalues.

The profiles of real and imaginary parts
of potentials (\ref{W1}) and (\ref{W2}) for $W_{0}=l=1$ are shown in Fig.~%
\ref{pt}(a), which shows that the harmonic-oscillator potential imposes
strong confinement on the wave function, even if imaginary potential (\ref%
{W1}) does not vanish at $|x|\rightarrow \infty $. Boundaries between
unbroken and broken $\mathcal{PT}$ symmetry, obtained by means of the
Fourier spectral method \cite{yjk,smm} for imaginary potentials $W_{1}(x)$
and $W_{2}(x)$ in the $(l,W_{0})$ plane, are plotted in Fig.~\ref{pt}(b).
The broken (unbroken) $\mathcal{PT}$-symmetry takes place above (under) the
boundaries. For example, if $l=2$ is fixed, the $\mathcal{PT}$
symmetry-breaking threshold for $W_{1}(x)$ and $W_{2}(x)$ is $W_{0}\approx
2.73$ and $2.46$, respectively. To further illustrate the $\mathcal{PT}$%
-symmetry breaking process for $W_{1}(x)$ and $W_{2}(x)$, eigenvalues of the
first four lowest energy levels are plotted as functions of $W_{0}$ for
fixed $l=2$ in Figs.~\ref{pt}(c1,c2,d1,d2). The plots demonstrate that the $%
\mathcal{PT}$ symmetry gets broken at critical points, through collisions of
the real eigenvalues corresponding to ground and first excited state at $%
W_{0}\approx 2.73$ for $W_{1}(x)$, or at $W_{0}\approx 2.46$ for $W_{2}(x)$.

\subsection{The linear-stability analysis}

To analyze the linear stability of stationary solutions (\ref{stationary}),
the perturbed solution of Eq.~(\ref{phipsi})
\begin{equation}
\begin{split}
\phi (x,z)& =\left\{ u(x)+\rho \left[ F_{1}(x)e^{-i\varepsilon
z}+G_{1}^{\ast }(x)e^{i\varepsilon ^{\ast }z}\right] \right\} e^{-i\mu z}, \\
\psi (x,z)& =\left\{ v(x)+\rho \left[ F_{2}(x)e^{-i\varepsilon
z}+G_{2}^{\ast }(x)e^{i\varepsilon ^{\ast }z}\right] \right\} e^{-i\mu z},
\end{split}
\label{perturbation}
\end{equation}%
are employed, where $0<\rho \ll 1$ is an infinitesimal amplitude of the
perturbations, $F_{j}(x)$ and $G_{j}(x)$ $(j=1,2)$ are eigenfunctions of the
linearized eigenvalue problem, and $\varepsilon $ is the respective
eigenvalue. Substituting the perturbed solution (\ref{perturbation}) into
the full nonlinear system of Eq.~(\ref{phipsi}) and linearizing around the
localized solution, we derive the corresponding Bogoliubov-de Gennes (BdG)
equations~\cite{Pit}:
\begin{equation}\label{m}
  \left(
    \begin{array}{cccc}
      L_{1} & -\lambda & L_3 & 0 \\
      -\lambda & L_2 & 0 & L_4 \\
      -L_3^* & 0 & -L_1 & \lambda \\
      0 & -L_4^* & \lambda & -L_2 \\
    \end{array}
  \right)\left(
           \begin{array}{c}
             F_1 \\
             F_2 \\
             G_1 \\
             G_2 \\
           \end{array}
         \right)=\varepsilon \left(
                               \begin{array}{c}
                                 F_1 \\
                                 F_2 \\
                                 G_1 \\
                                 G_2 \\
                               \end{array}
                             \right),
\end{equation}
\begin{equation}
L_{1}=-\frac{1}{2}\partial _{x}^{2}-2|u|^{2}+\frac{1}{2}x^{2}+iW(x)-\mu ,\quad
L_{2}=-\frac{1}{2}\partial _{x}^{2}-2|v|^{2}+\frac{1}{2}x^{2}+iW(x)-\mu , \quad
L_{3}=-u^{2},\quad L_{4}=-v^{2}.
\end{equation}%
The stationary states are linearly unstable if at least one eigenvalue $%
\varepsilon $ is complex. To clearly identify the change of spectra of the
above eigenvalue problem produced by the numerical solution, we set the
linear-instability index
\begin{equation}
\tau =\log _{10}(\max \{|\mathrm{Im}(\varepsilon )|\}),  \label{index}
\end{equation}%
and categorize the underlying solution as a stable one if $\tau \leq -8$,
otherwise the solution is unstable.

\begin{figure}[!t]
\centering
{\scalebox{0.6}[0.6]{\includegraphics{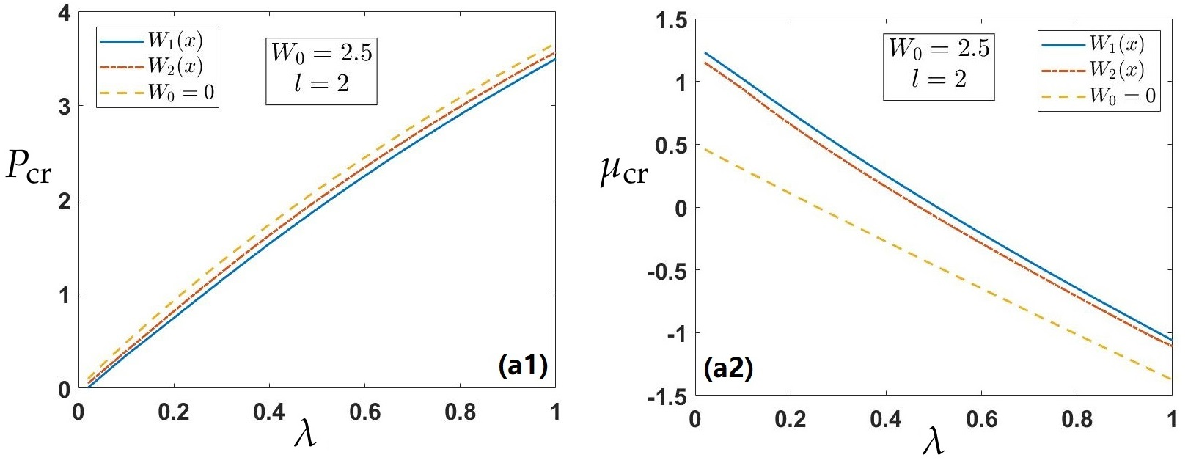}}}
\vspace{0.1in}
\caption{{\protect\small
Boundaries of the $\mathcal{PT}$ symmetry breaking for the
inter-core symmetry $P_{\mathrm{cr}}(\lambda)$ and  propagation constant $-\mu _{\mathrm{cr}}$.
(a1) The boundary for the breaking of the
inter-core symmetry, $P_{\mathrm{cr}}(\protect\lambda )$, for the }$\mathcal{%
PT}${\protect\small -symmetric potentials $W_{1}(x)$ and $W_{2}(x)$ at $%
W_{0}=2.5$ and $l=2$, respectively. (a2) Critical values of propagation
constant $-\protect\mu _{\mathrm{cr}}$ at the inter-core symmetry-breaking
point of the GS modes versus the inter-core coupliing constant $\protect%
\lambda $ for the potentials $W_{1}(x)$ and $W_{2}(x)$ at $W_{0}=2.5$ and $%
l=2$, respectively. The yellow dotted line shows the results for $W_{0}=0$,
i.e., for the pure real potential. }}
\label{GS_ssb}
\end{figure}

\section{Symmetry breaking of nonlinear modes and their stability}

In this section, we concentrate on the SBB of nonlinear modes including
ground-state and dipole modes and their stability. Especially, the influence
of the imaginary part $W(x)$ of the $\mathcal{PT}$-symmetric potential on
SBB is investigated. The modified squared-operator iteration method \cite%
{yjk,msom} is used to find the numerical solutions. And based on the Fourier
collocation method, the stability of these solutions is discussed by solving
the BdG equations (\ref{m}). Finally, the dynamics of symmetric and
asymmetric modes is further studied by direct simulations of the perturbed
solution.

\begin{figure}[!t]
\centering
{\scalebox{0.86}[0.86]{\includegraphics{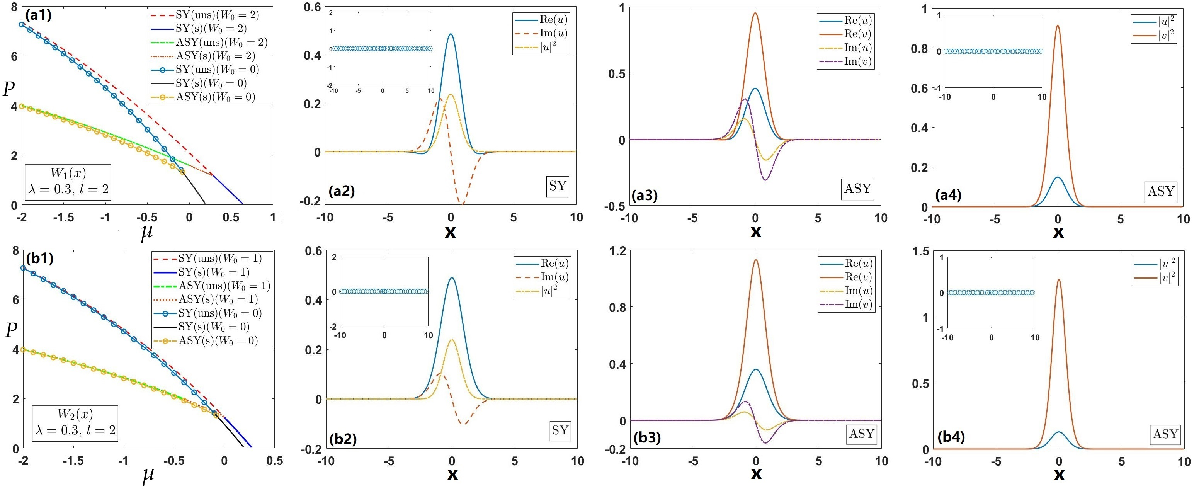}}}\hspace{-0.35in} \vspace{%
0.15in}
\caption{{\protect\small  Total power $P$ versus propagation constant $-%
\protect\mu $ for symmetric (SY) and asymmetric (ASY) GS modes. (a1) The }$%
\mathcal{PT}${\protect\small -symmetric potential is $W_{1}(x)$, with $%
W_{0}=2$, $l=2$ and inter-core coupling constant }$\protect\lambda =0.3$.%
{\protect\small \ (b1) Potential $W_{2}(x)$ with $W_{0}=1$, $l=2$, and $%
\protect\lambda =0.3$, The results corresponding to the real potential (with
$W_{0}=0$) are also presented. (a2,b2) Profiles of stable symmetric GS
solutions $u(x)$ for $W_{1}(x)$, $\protect\mu =0.4$, and $W_{2}(x)$, $%
\protect\mu =0.1$, respectively. (a3,a4,b3,b4) Profiles of stable asymmetric
GS solutions $u(x)$ and $v(x)$ for $W_{1}(x)$, $\protect\mu =0.1$ and $%
W_{2}(x)$, $\protect\mu =-0.4$, respectively. Here \textquotedblleft uns"
and \textquotedblleft s" refer to unstable and stable solutions,
respectively. The inset shows stability spectra for eigenvalues $\protect%
\varepsilon $ of small perturbations, see Eq. (\protect\ref{perturbation}).}}
\label{mup12}
\end{figure}

\subsection{Symmetric and asymmetric ground-state (GS) modes}

\begin{figure}[!t]
\centering
{\scalebox{0.75}[0.65]{\includegraphics{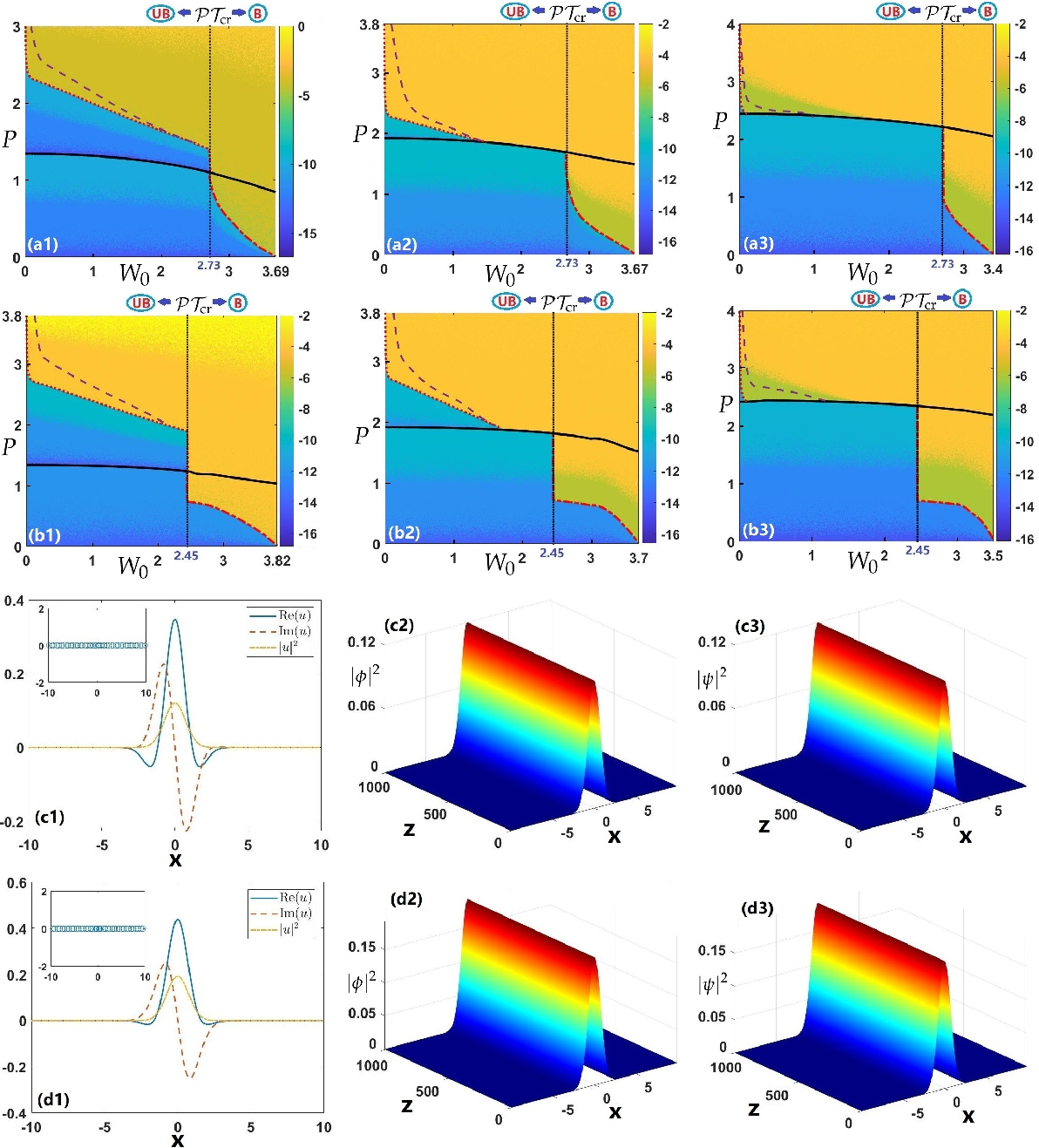}}}  \hspace{-0.4in}
\vspace{0.1in}
\caption{{\protect\small The stability and boundary of the breaking of the
inter-core symmetry for symmetric and asymmetric GS modes in $(W_{0},P)$-space with $l=2$. The colorbar denotes the linear instability index $\protect\tau $ defined in Eq.~(\protect\ref{index}). Panels (a1,a2,a3)
correspond to symmetric and asymmetric GS modes for the} $\mathcal{PT}$%
{\protect\small -symmetric potential $W_{1}(x)$ and inter-core coupling
constants $\protect\lambda =0.3$, $0.45$ and $0.6$, respectively. Panels
(b1,b2,b3) correspond to symmetric and asymmetric GS modes for $W_{2}(x)$
and $\protect\lambda =0.3$, $0.45$ and $0.6$, respectively. Here the black
solid line represents the boundary for the breaking of the inter-core
symmetry, $P_{\mathrm{cr}}(W_{0})$. Regions below and above the black line
represent, respectively, symmetric (SY) solutions and asymmetric ones (ASY).
The red dotted line represents the instability boundary of symmetric
solutions (asymmetric solutions) below (above) the black line. The yellow
instability area of asymmetric solutions above the purple dotted line
represents the instability (collapse) shown in Fig.~\protect\ref{GS_evo}%
(b1-b4), while the oscillating instability displayed in Fig.~\protect\ref%
{GS_evo}(c1-c4) takes place in the yellow area below the purple dotted line.
The vertical dotted line indicates the $\mathcal{PT}$-symmetry breaking
point ($\mathcal{PT}_{\mathrm{cr}}\approx 2.73$ for potential $W_{1}(x)$,
and $\mathcal{PT}_{\mathrm{cr}}\approx 2.45$ for $W_{2}(x)$), where
\textquotedblleft UB" (\textquotedblleft B") denotes the phases with
unbroken (broken) $\mathcal{PT}$ symmetry. (c1,c2,c3) Profiles of stable
symmetric GS solutions $u(x)$ for $\protect\lambda =0.3$ and $W_{1}(x)$ with
$W_{0}=2.8$, $l=2$ ($\mathcal{PT}$ breaking) and $\protect\mu =1$ ($%
P\approx 0.437$), and the corresponding numerically simulated evolution.
(d1,d2,d3) Profiles of stable symmetric GS solutions $u(x)$ for $\protect%
\lambda =0.3$ and $W_{2}(x)$ with $W_{0}=2.5$, $l=2$ ($\mathcal{PT}$ breaking) and $\protect\mu =0.6$ ($P\approx 0.7109$) and the
corresponding numerically simulated evolution. }}
\label{GS_w0p}
\end{figure}

\begin{figure}[!t]
\centering
{\scalebox{0.84}[0.84]{\includegraphics{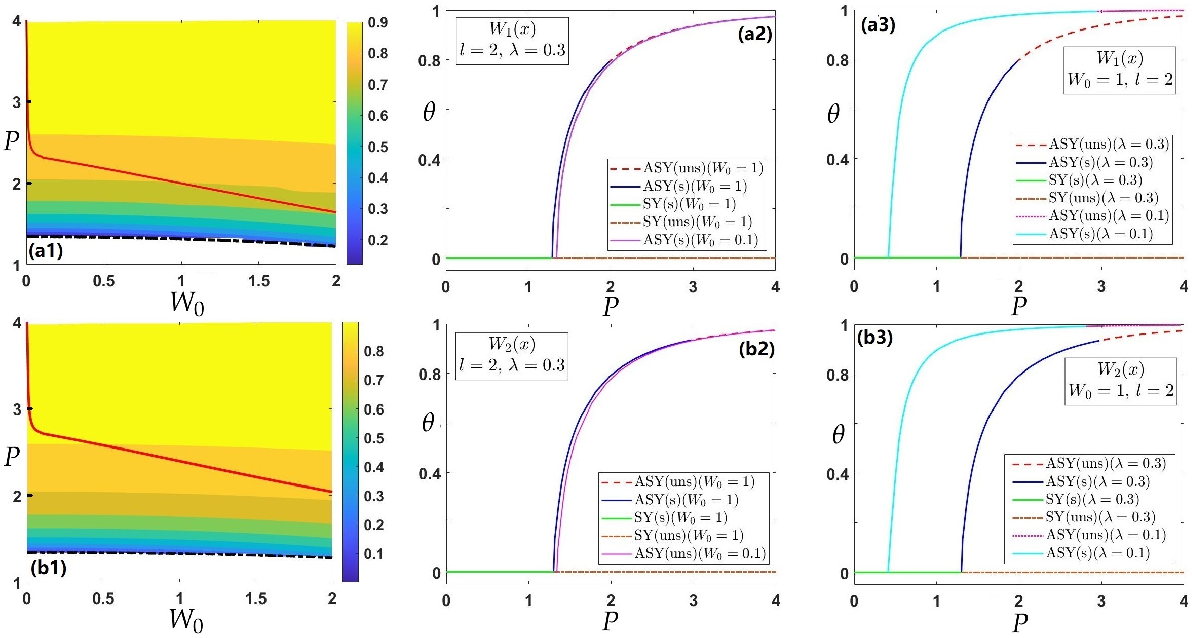}}}\hspace{-0.35in}
\vspace{0.1in}
\caption{{\protect\small (a1,b1) Asymmetry characteristic $\protect\theta $
(see Eq.~(\protect\ref{asym})) in $(W_{0}, P)$-space for the }$%
\mathcal{PT}${\protect\small -symmetric potentials $W_{1}(x)$ and $W_{2}(x)$
with $l=2$, respectively, and $\protect\lambda =0.3$. Here the red line is
the stability boundary for the asymmetric states. Bifurcation diagrams in
the $(P,\protect\theta )$ plane for the GS modes with different parameters:
(a2,b2) the bifurcation diagrams for potentials $W_{1}(x)$ and $W_{2}(x)$
with $l=2$ and $\protect\lambda =0.3$; (a3,b3) the bifurcation diagrams for $%
W_{1}(x)$ and $W_{2}(x)$ with $W_{0}=1$ and $l=2$. Here \textquotedblleft
ASY(uns)" and \textquotedblleft ASY(s)" represent unstable and stable
asymmetric solutions, respectively, while \textquotedblleft SY(uns)" and
\textquotedblleft SY(s)" represent unstable and stable symmetric solutions. }
}
\label{Ptheta}
\end{figure}

At first, for GS solutions, the boundary for the breaking of the inter-core
symmetry $P_{\mathrm{cr}}(\lambda )$ can be obtained for the imaginary
potentials $W_{1}(x)$ and $W_{2}(x)$ at fixed potential parameters, \textit{%
viz}., $W_{0}=2.5$ and $l=2$, see Fig.~\ref{GS_ssb}(a1). For a fixed
coupling constant $\lambda $, when the total power $P$ exceeds the threshold
$P_{\mathrm{cr}}$, the symmetry breaking occurs and asymmetric solutions
appear. With the increase of $\lambda $, larger values of power $P$ are
naturally required for the occurrence of the symmetry breaking. The power
required for symmetry breaking under the action of potential $W_{2}(x)$ is
slightly larger than that required in the case of $W_{1}(x)$. Meanwhile,
when $W_{0}=0$ (i.e., the potential is real), the boundary for the breaking
of the inter-core symmetry is also exhibited in Fig.~\ref{GS_ssb}(a1). By
comparison, we see that a larger gain-loss strength reduces the power needed
for the symmetry breaking, but not significantly. Furthermore, Fig.~\ref%
{GS_ssb}(a2) displays the critical values of propagation constant $-\mu _{%
\mathrm{cr}}$ at the symmetry-breaking point of the GS modes for imaginary
potentials $W_{1}(x)$ and $W_{2}(x)$, respectively. According to the results
obtained for the real potential (see the yellow dotted line in Fig.~\ref%
{GS_ssb}(a2)), the effect of the imaginary part $W(x)$ of the potential on
the relation between the propagation constant $-\mu $ and power $P$ is
obvious.

To clearly display the relation between $\mu $ and $P$, the results for the
symmetric and asymmetric GS modes are presented by means of $P(\mu )$
dependences for potential $W_{1}(x)$ at $W_{0}=2$ and $W_{2}(x)$ at $W_{0}=1$
in Figs.~\ref{mup12}(a1,b1), respectively. The results for the real
potential, with $W_{0}=0$, are also exhibited for the comparison. It is seen
that, with the increase of power $P$, the curves for the real and complex $%
\mathcal{PT}$-symmetric potentials gradually coalesce. Note that all the
results satisfy the Vakhitov-Kolokolov (VK) criterion ($\mathrm{d}P/\mathrm{d%
}\mu <0$), which is the necessary condition for stability of solutions in
the case of any self-focusing nonlinearity (see the original work \cite{vk}
and detailed explanations in Refs. \cite{Berge,Sulem,Fibich}). However, the
asymmetric GS modes are only partially stable. For instance, for potential $%
W_{1}(x)$ with $W_{0}=2$ and $l=2$, the asymmetric GS becomes unstable at $%
P>1.5878$. This is in contrast with the case of the real potential, in which
case the asymmetric GS is always stable when it exists \cite{Viskol}. In the
limit of $\mu \rightarrow -\infty $, the small component of an asymmetric
solution is $v(x)\approx -(\lambda /\mu )u(x)$ according to Eq.~(\ref{uv}).
Therefore, for fixed coupling constant $\lambda $, $P_{2}=\int_{-\infty
}^{+\infty }|\psi (x,z)|^{2}dx\rightarrow 0$ as $\mu \rightarrow -\infty $.
Thus, in the limit of $\mu \rightarrow -\infty $, the total powers of the
symmetric and asymmetric solutions are related so that
\bee
P_{\mathrm{sym}}\approx 2P_{\mathrm{asym}}.
\ene
In Figs.~\ref{mup12}(a2-a4,b2-b4), some
typical examples of the stable symmetric and asymmetric GS solutions for
imaginary potentials $W_{1}(x)$ and $W_{2}(x)$ are exhibited.

\begin{figure}[!t]
\centering
{\scalebox{0.82}[0.82]{\includegraphics{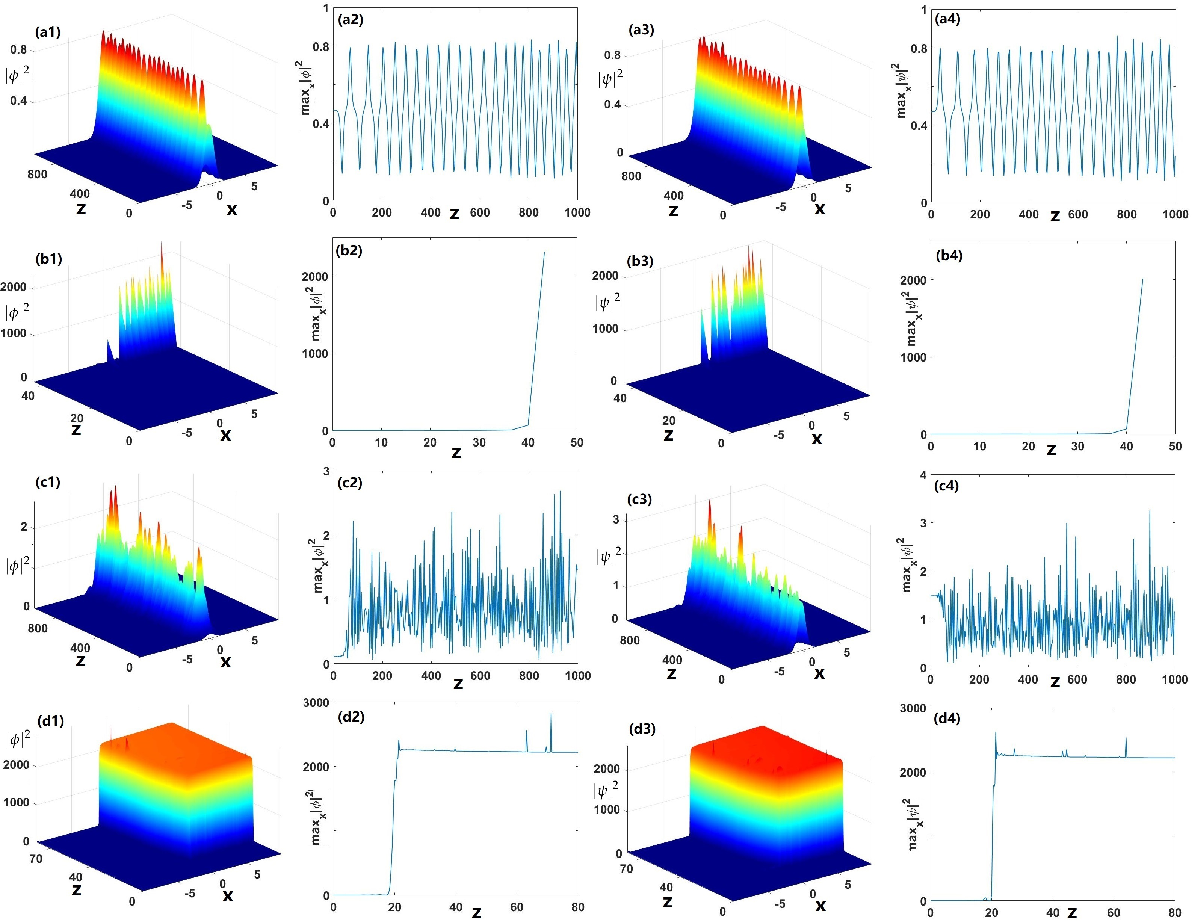}}}  \hspace{-0.4in} \vspace{0.15in}
\caption{{\protect\small The numerically simulated evolution of an unstable
GS state and the respective evolution of maximum densities of the components
in the parallel cores for the }$\mathcal{PT}${\protect\small -symmetric
potential $W_{1}(x)$ and inter-core coupling constant $\protect\lambda =0.3$%
. (a1-a4) A symmetric GS solution for $W_{0}=1$ and $l=2$ at $\protect\mu %
=-0.045$ ($P\approx 1.5$), which shows the oscillatory instability. (b1-b4)
A symmetric GS solution for $W_{0}=2.8$ and $l=2$ at $\protect\mu =0.75$ ($%
P\approx 0.9$), which shows the collapse of the soliton. (c1-c4) An
asymmetric GS solution with $W_{0}=1$ and $l=2$ at $\protect\mu =-0.5$ ($%
P\approx 2.14$), which shows the oscillatory instability. (d1-d4) An
asymmetric GS solution with $W_{0}=2.8$ and $l=2$ at $\protect\mu =0.4$ ($%
P\approx 1.32$), which shows the collapse. }}
\label{GS_evo}
\end{figure}

Note that, for the GS modes, the results produced by the imaginary
potentials $W_{1}(x)$ and $W_{2}(x)$ are similar when their amplitude $W_{0}$
is small. However, when the strength of gain-loss distribution is large, the
corresponding results for the SBB and stability are different. Figure~\ref%
{GS_w0p} exhibits the stability and boundary of the breaking of the
inter-core symmetry for symmetric and asymmetric GS modes in the plane of $%
(W_{0},P)$. It is observed that the imaginary part $W(x)$ of the potential
produces an effect on the breaking of the inter-core symmetry and stability.
With the increase of $W_{0}$, the critical values of the power at
symmetry-breaking point gradually decrease. Furthermore, for conservative
systems the boundary of the inter-core symmetry breaking is, generally, a
boundary of destabilization of the symmetric GS modes. However, near the $%
\mathcal{PT}$-symmetry breaking point, the stability of the symmetric
solution becomes unstable at smaller $P$. For example, for the potential $%
W_{1}(x)$ with $l=2$, the $\mathcal{PT}$-symmetric breaking occurs at $%
W_{0}>2.73$, according to Fig.~\ref{pt}(b), which leads to the onset of
instability of symmetric solutions before the inter-core symmetry breaking
occurs (see Figs.~\ref{GS_w0p}(a1,a2,a3) corresponding to $\lambda =0.3,0.45$
and $0.6$, respectively). On the other hand, an important finding is that
there still exist stable symmetric solutions in the region of the broken $%
\mathcal{PT}$ symmetry. For example, for the potential $W_{1}(x)$ with $%
W_{0}=2.8$ and $l=2$, stable symmetric solutions can be found at $P\approx
0.437$ and $\lambda =0.3$ as is shown in Figs.~\ref{GS_w0p}(c1,c2,c3). For
the potential $W_{2}(x)$ with $W_{0}=2.5$ and $l=2$, stable symmetric
solutions are obtained at $P\approx 0.7109$ and $\lambda =0.3$, see Figs.~%
\ref{GS_w0p}(d1,d2,d3). When $W_{0}=0$, i.e., the potential is real, the
asymmetric GS modes, are always stable. However, when the imaginary part $%
W(x)$ of the potential is added, the asymmetric solutions becomes unstable
when power $P$ exceeds a certain threshold, as shown by the red dotted line
above the black line in Fig.~\ref{GS_w0p}. As $W_{0}$ increases, the
stability region of the asymmetric solution shrinks. Eventually, there is no
stable asymmetric solution after the $\mathcal{PT}$-symmetry breaking
occurs. When the inter-core coupling constant $\lambda $ is larger, the
critical power at the point of the breaking of the inter-core symmetry is
larger too. As $\lambda $ increases, the stability region of the asymmetric
solutions shrinks. For example, there is a conspicuous stability region for
the asymmetric states at $\lambda =0.45$, which almost disappears at $%
\lambda =0.6$, see Figs.~\ref{GS_w0p}(a3,b3).

The SBB of the GS modes is illustrated by heatmaps of asymmetry parameter $%
\theta $, see Eq.~(\ref{asym}), in the plane of $(W_{0},P)$ for imaginary
potentials $W_{1}(x)$ and $W_{2}(x)$ with $\lambda =0.3$ and $l=2$ in Figs.~%
\ref{Ptheta}(a1,b1). The bifurcation diagrams are also exhibited in Figs.~%
\ref{Ptheta}(a2,a3,b2,b3) by means of $\theta (P)$ curves for fixed
parameters of the system. These diagrams clearly demonstrate that the SBB
for the GS modes is of the supercritical type, i.e., the emerging branches
of the asymmetric solutions go forward (rather than backward) as the power
keeps growing past the SBB\ point, according to the usual definition \cite%
{bif}. On the other hand, strength $W_{0}$ of the gain-loss distribution has
a little effect on $\theta $, while it affects the stability of the
solutions. It is also worthy to note that the stability region is much
larger for the\ $\mathcal{PT}$ potential $W_{2}(x)$ than for $W_{1}(x)$.

\begin{figure}[!t]
\centering
{\scalebox{0.86}[0.86]{\includegraphics{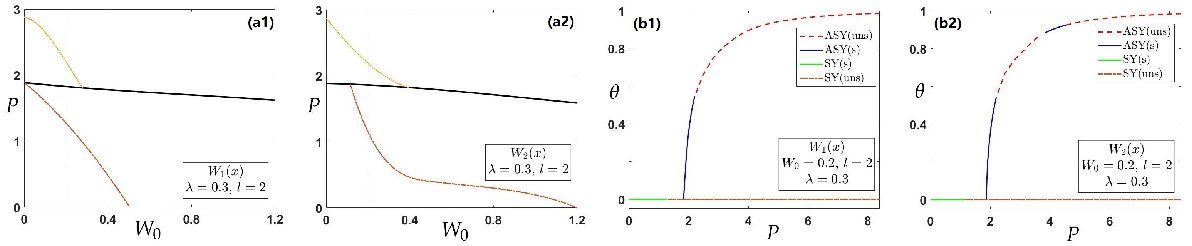}}} \hspace{-0.4in}
 \vspace{0.1in}
\caption{{\protect\small The stability and the boundary of the breaking of
the inter-core symmetry for symmetric and asymmetric DM modes in $(W_{0},P)$-space with $\protect\lambda =0.3$, $l=2$ and imaginary potentials $%
W_{1}(x)$ in (a1) and $W_{2}(x)$ in (a2). Here the black solid line represents
the boundary of the breaking of the inter-core symmetry. The regions below
and above the black line are populated, respectively, by the symmetric (SY)
and asymmetric (ASY) DM solutions. The red (yellow) dotted line represents
the instability boundary of symmetric (asymmetric) solutions below (above)
the black line. The region below the dotted line represents stable
solutions. (b1,b2) Bifurcation diagrams in the $(P,\protect\theta )$ plane
for imaginary potentials $W_{1}(x)$ and $W_{2}(x)$ with $W_{0}=0.2$, $l=2$
and $\protect\lambda =0.3$. Here \textquotedblleft ASY(uns)" and
\textquotedblleft ASY(s)" represent unstable and stable asymmetric
solutions, respectively. \textquotedblleft SY(uns)" and \textquotedblleft
SY(s)" represent unstable and stable symmetric solutions, respectively. }}
\label{ssbj}
\end{figure}

\subsection{Dynamics of symmetric and asymmetric ground-state (GS) modes}

Figure~\ref{GS_evo} shows the numerically simulated evolution of unstable GS
states, to which random perturbation are added at the 2\% amplitude level,
in the case of the $\mathcal{PT}$-symmetric potential $W_{1}(x)$ and
inter-core coupling constant $\lambda =0.3$. For symmetric GS modes, the SBB
causes oscillatory instability, see Figs.~\ref{GS_evo}(a1-a4). In the
simulations, the symmetric state spontaneously transforms into an asymmetric
one with residual oscillations. Figures~\ref{GS_evo}(a2, a4) demonstrate
that minimum and maximum amplitudes of both components are attained at the
same time. However, in the $\mathcal{PT}$-broken region the symmetric GS is
collapses (see Figs.~\ref{GS_evo}(b1-b4)), where the amplitude of
the imaginary part of the potential is $W_{0}=2.8$.

Asymmetric GS modes become unstable when their power $P$ exceeds a certain
threshold value, as shown in Fig.~\ref{GS_w0p}. At first, the asymmetric
solutions exhibit oscillating instability (see Figs.~\ref{GS_evo}(c1-c4)),
which is followed by the onset of collapse with the increase of $P$. The
corresponding instability boundary is displayed by the purple dotted line in
Fig.~\ref{GS_w0p}. The yellow instability area of asymmetric solutions above
the purple dotted line represents the collapse, while the oscillatory
instability takes place in the yellow area below the purple dotted line.
Simultaneously, due to the presence of $\mathcal{PT}$-symmetry breaking, the asymmetric GS
solutions start to collapse.

\subsection{Symmetric and asymmetric dipole modes (DM) and their dynamics}

Similarly, systematic results for symmetric and asymmetric DMs (first excited states) have been
collected too. First, the critical values of power $P_{\mathrm{cr}}(W_{0})$
(the black solid line) at the symmetry-breaking point for the imaginary
potentials $W_{1}(x)$ and $W_{2}(x)$, respectively, are exhibited in Figs.~%
\ref{ssbj}(a1) and (a2). At $P>P_{\mathrm{cr}}$, the inter-core symmetry
breaking occurs and symmetric DMs appear. Compared to the GS modes, the
critical value of the power is somewhat higher for the dipoles than for the
GS modes, cf. Figs.~\ref{ssbj}(a1,a2) and Figs.~\ref{GS_w0p}(a1,b1). This
happens because the shape of the dipoles is broader than that of GS modes,
which makes the nonlinearity weaker for the dipoles, in comparison with the
GS mode. Besides, the stable region of symmetric DMs shrinks rapidly as $%
W_{0}$ increases, especially for the $W_{1}(x)$ potential, and this
instability penetrates into a part of the originally stable branches of the
symmetric DMs below the SBB point, as shown in Figs.~\ref{ssbj}(a1,a2),
where the red dotted line represents the instability boundary of symmetric
solutions below the black line.

Fixing $W_{0}=0.2$, $l=2$ and $\lambda =0.3$, the bifurcation diagrams in
the $(P,\theta )$ plane for imaginary potentials $W_{1}(x)$ and $W_{2}(x)$
are shown, severally, in Figs.~\ref{ssbj}(b1) and (b2). The diagrams clearly
demonstrate that the SBB for the DMs is of the supercritical type. Under the
action of the imaginary part of the potential, asymmetric DMs feature
partial stability, with respective complex instability eigenvalues. The
instability boundary of asymmetric solutions is displayed by yellow dotted
line in Figs.~\ref{ssbj}(a1,a2).

\begin{figure}[t]
\centering
{\scalebox{0.85}[0.85]{\includegraphics{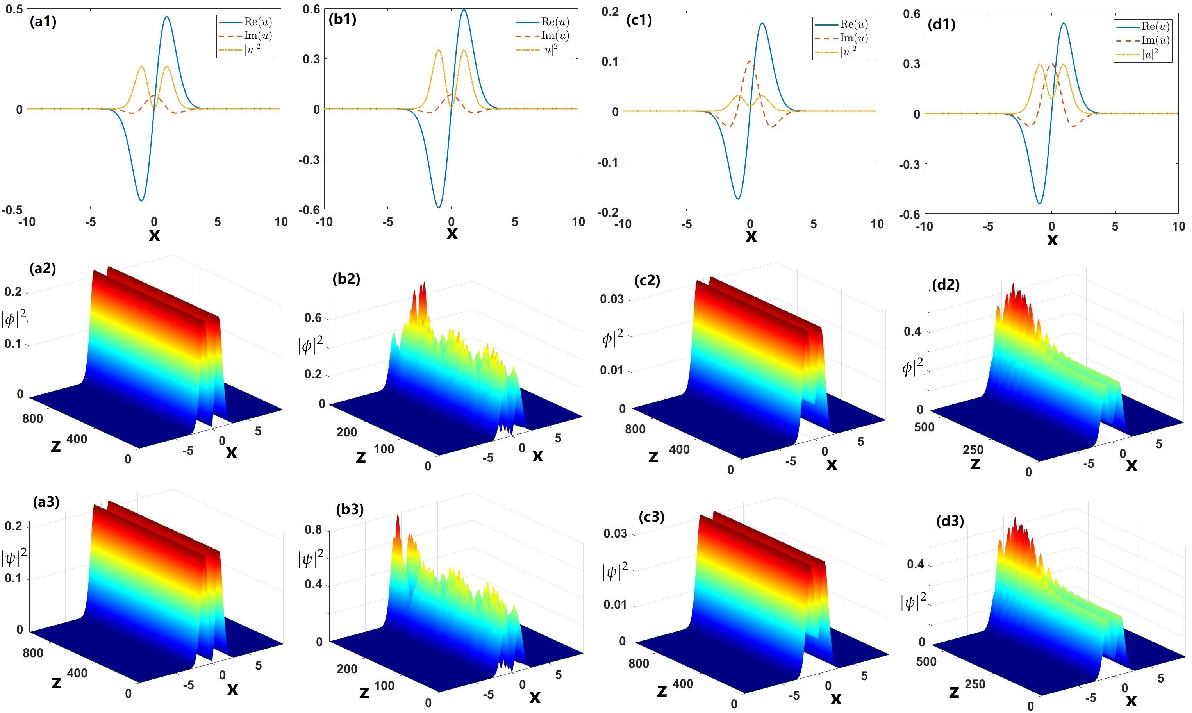}}}  \hspace{-0.4in}
 \vspace{0.15in}
\caption{{\protect\small Profiles of stable and unstable symmetric DM
solutions $u(x)$ at $\protect\lambda =0.3$ for imaginary potentials $%
W_{1}(x) $ and $W_{2}(x)$ with $l=2$, and the corresponding numerically
simulated evolution. Panels (a1,a2,a3) display stable DM solutions for $%
W_{1}(x)$ with $W_{0}=0.2$ at $\protect\mu =1.05$ ($P\approx 1.0012$).
Panels (b1,b2, b3) display unstable DM solutions for $W_{1}(x)$ with $%
W_{0}=0.2$ at $\protect\mu =0.95$ ($P\approx 1.6399$). Panels (c1,c2,c3)
display stable DM solutions for $W_{2}(x)$ with $W_{0}=1$ at $\protect\mu %
=1.1$ ($P\approx 0.165$). Panels (d1,d2,d3) display unstable DM solutions
for $W_{2}(x)$ with $W_{0}=1$ at $\protect\mu =0.98$ ($P\approx 1.5158$). }}
\label{jsym}
\end{figure}

\begin{figure}[!t]
\centering
{\scalebox{0.85}[0.8]{\includegraphics{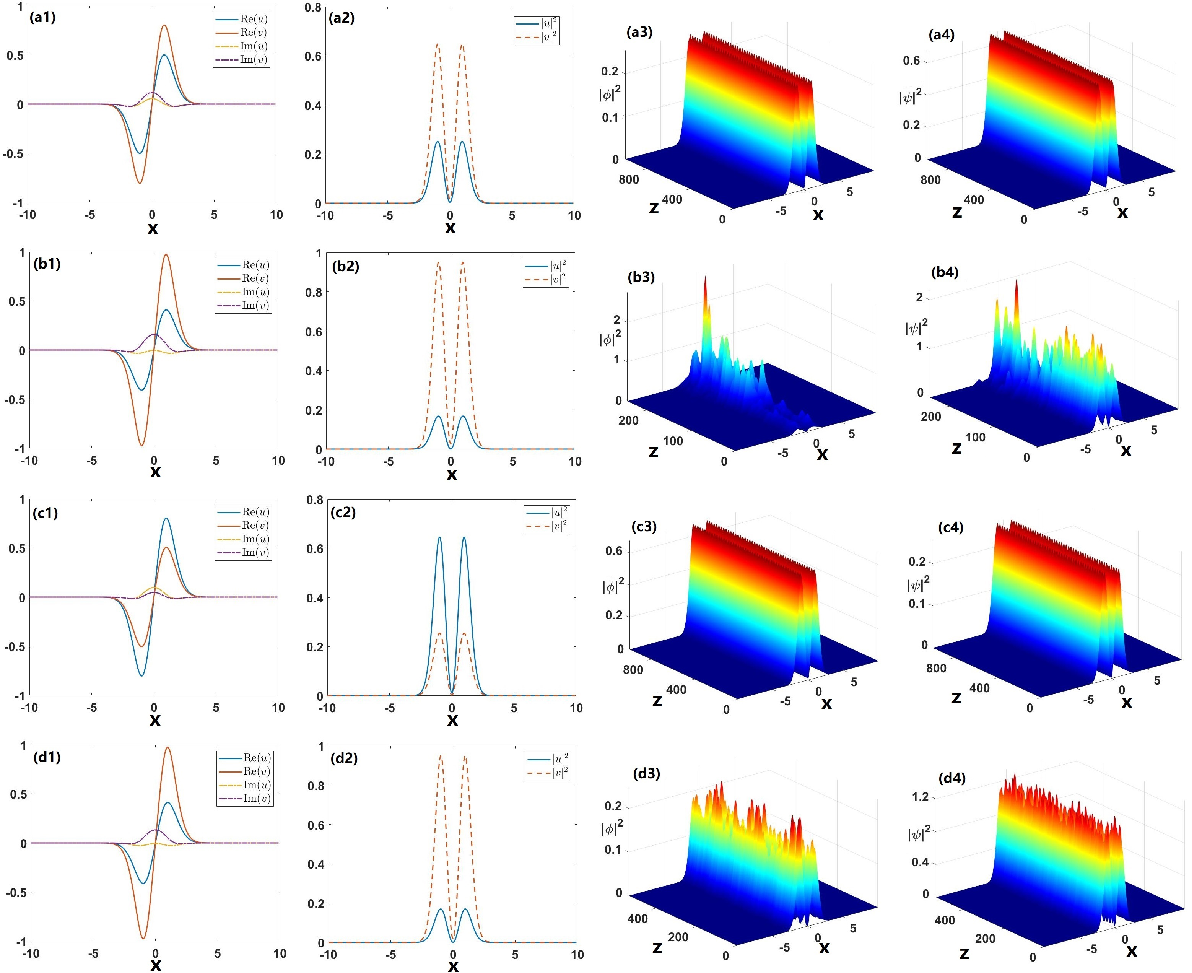}}}  \hspace{-0.4in}
\vspace{0.15in}
\caption{{\protect\small Profiles of stable and unstable asymmetric DM
solutions $u(x)$ and $v(x)$ at $\protect\lambda =0.3$ for imaginary
potentials $W_{1}(x)$ and $W_{2}(x)$ with $W_{0}=0.2$ and $l=2$, and the
corresponding numerically simulated evolution. (a1-a4): Stable DM solutions
for $W_{1}(x)$ at $\protect\mu =0.85$ ($P\approx 2.0577$). (b1-b4): Unstable
DM solutions for $W_{1}(x)$ at $\protect\mu =0.7$ ($P\approx 2.4825$).
(c1-c4): Stable DM solutions for $W_{2}(x)$ at $\protect\mu =0.85$ ($%
P\approx 2.0558$). (d1-d4): Unstable DM solutions for $W_{1}(x)$ at $\protect%
\mu =0.7$ ($P\approx 2.4818$). }}
\label{jasym}
\end{figure}

Figures~\ref{jsym} and \ref{jasym} display profiles of symmetric and
asymmetric DM solutions, as well as the corresponding numerically simulated
evolution with random noise at the 2\% amplitude level. In general, when the
strength of the gain-loss distribution is weak, the DM solutions exhibit
oscillatory instability, as shown in the Figs.~\ref{jsym}(b2,b3). However,
at larger $W_{0}$ the DMs tend to collapse, see Figs.~\ref{jsym}(d2,d3).

\section{Symmetric and asymmetric GS modes for the pure imaginary
localized potential}

\begin{figure}[!t]
\centering
{\scalebox{0.84}[0.84]{\includegraphics{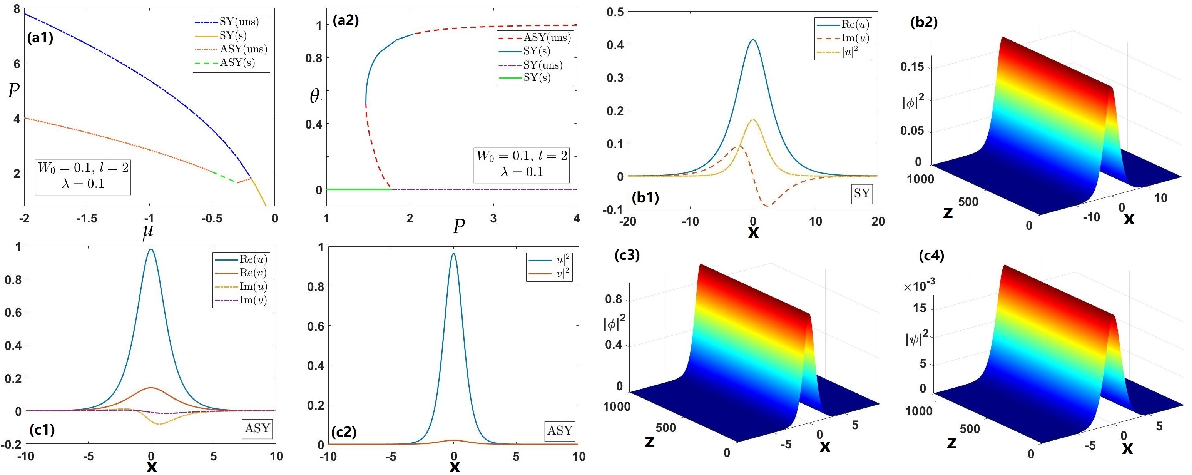}}} \hspace{-0.4in} \vspace{0.15in}
\caption{{\protect\small (a1) Total power $P$ versus propagation constant $%
\protect-\mu $ for symmetric (SY) and asymmetric (ASY) GS modes in the case
of the pure imaginary localized potential $W_{2}(x)$. (a2) Bifurcation
diagrams in the $(P,\protect\theta )$ plane for the imaginary potential $%
W_{2}(x)$ with $W_{0}=0.1$, $l=2$, and inter-core coupling constant $\protect%
\lambda =0.1$. (b1,b2). The stable symmetric GS solution $u(x)=v(x)$ at $%
\protect\mu =-0.18$ and its evolution. (c1-c4). Profiles $u(x)$ and $v(x)$
of the stable asymmetric GS solution at $\protect\mu =-0.5$ and its
evolution. Here \textquotedblleft uns" and \textquotedblleft s" represent,
respectively, unstable and stable solutions, and the inter-core coupling
constant is $\protect\lambda =0.1$. }}
\label{v0}
\end{figure}

It is relevant to consider the special case of system (\ref{phipsi}) with $%
V(x)=0$ and a purely imaginary localized imaginary potential $W_{2}(x)$ (in
this case, the delocalized imaginary potential $W_{1}(x)$ would definitely
make everything unstable in the case of $V(x)=0$). For the gain-loss
distribution represented by the imaginary potential with a small strength,
we set $W_{0}=0.1$, $l=2$, while the coupling constant is fixed as $\lambda
=0.1$. A typical $P(\mu )$ dependence for the symmetric and asymmetric GS
modes obtained in this case is displayed in Fig.~\ref{v0}(a1), being stable
in narrow intervals. In particular, the symmetric GS modes are stable in the
range of $1.554<P<1.8122$. For example, when $\mu =-0.18$ $(P\approx 1.736)$%
, the profile of the stable symmetric GS solution $u(x)=v(x)$ and its stable
evolution are shown in Figs.~\ref{v0}(b1,b2). As the power decreases, the
width of the solitons increases and their amplitude decreases. Furthermore,
when $\mu <-0.19$, asymmetric GS solutions bifurcate from the symmetric
branch. In a narrow interval of $\mu \in (-0.3,-0.19)$, its power slightly
decreases, i.e., $\mathrm{d}P/\mathrm{d}\mu >0$, hence they not satisfy the
above-mentioned VK criterion. In interval $\mu \in \lbrack -0.5,-0.3]$, the
power starts to grow ($\mathrm{d}P/\mathrm{d}\mu <0$) and the asymmetric GS
solutions are found. For example, when $\mu =-0.5$ $(P\approx 2.0374)$, the
profiles of the stable asymmetric GS solution, $u(x)$ and $v(x)$, as well as
its stable evolution are exhibited in Figs.~\ref{v0}(c1,c2,c3,c4). The
corresponding bifurcation diagram is exhibited in Fig.~\ref{v0}(a2). It
shows that the type of the SBB is subcritical, similar to the weakly
subcritical one for the free-space 1D solitons in the linearly coupled NLS
equations. For larger values of the strength of gain-loss distribution $W_{0}
$, symmetric and asymmetric GS solutions can still be found, but they are
unstable.

\section{Management of symmetric and asymmetric GS modes via the
variable $\mathcal{PT}$-symmetric potential or inter-core coupling}

Finally, we aim to discuss dynamical transformation (\textquotedblleft
management") of symmetric and asymmetric GS modes by making the potential
parameter $W_{0}$ or coupling constant $\lambda $ functions of propagation
distance $z$, i.e., we set $W_{0}=\widehat W_{0}(z)$ in Eqs.~(\ref{W1}) and (\ref{W2}),
and $\lambda =\widehat \lambda (z)$ in Eq. (\ref{phipsi}). The accordingly modified
system is
\begin{equation}
\begin{split}
i\phi _{z}& =-\frac{1}{2}\phi _{xx}-|\phi |^{2}\phi +(V(x)+i\widehat W(x,z))\phi
-\widehat \lambda (z)\psi , \\
i\psi _{z}& =-\frac{1}{2}\psi _{xx}-|\psi |^{2}\psi +(V(x)+i\widehat W(x,z))\psi
-\widehat \lambda (z)\phi .
\end{split}
\label{phipsij}
\end{equation}
where  $V(x)=\frac12 x^2$, and $\widehat W(x,z)$ is chosen as
\begin{equation}
\widehat W_1(x,z)=\widehat W_{0}(z)\tanh \left( \frac{x}{l}\right),  \label{W1g}
\end{equation}%
or
\begin{equation}
\widehat W_2(x,z)=\widehat W_{0}(z)\,\mathrm{sech}\left( \frac{x}{l}\right) \tanh \left( \frac{x}{%
l}\right) ,  \label{W2g}
\end{equation}
Here we address two scenarios of the adiabatic variation of parameters $%
\widehat W_{0}(z)$ and $\widehat \lambda (z)$ in the form [cf. Ref.~\cite{yan15,song22a,song22b}]:
\begin{equation}
\chi (z)\!=\!\left\{ \!%
\begin{array}{ll}
\displaystyle\frac{\chi_{2}-\chi_{1}}{2}\left[ 1-\cos \left( \frac{%
10\pi z}{z_{\mathrm{max}}}\right) \right] +\chi_{1}, & \displaystyle%
0\leq z<\frac{z_{\mathrm{max}}}{10}, \v \\
\chi_{2}, & \displaystyle\frac{z_{\mathrm{max}}}{10}\leq z\leq z_{%
\mathrm{max}},%
\end{array}%
\right.  \label{lambda}
\end{equation}%
where $\chi_{1,2}$ stand for the initial and final values in the
excitation, respectively, and $z_{\mathrm{max}}$ represents the maximal
value of numerical simulations.

\begin{figure}[!t]
\centering
{\scalebox{0.84}[0.84]{\includegraphics{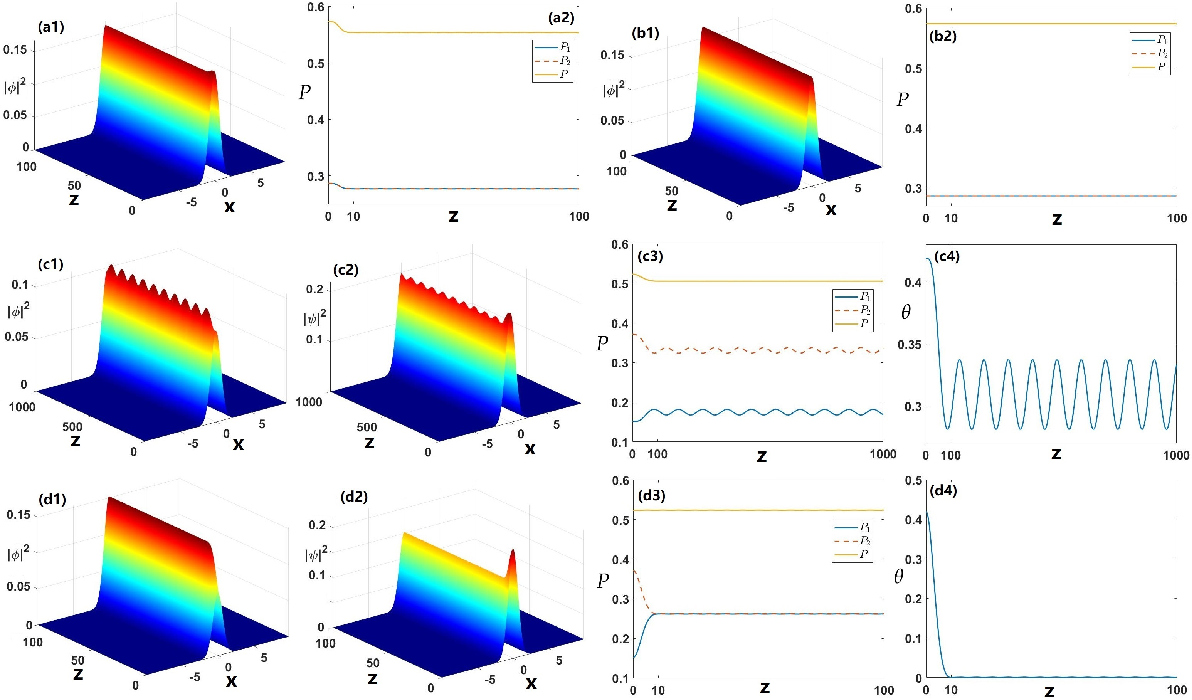}}} \hspace{-0.4in}\vspace{0.15in}
\caption{{\protect\small Transformation of GS modes in the framework of Eq. (%
\protect\ref{lambda}). The initial state corresponds to $W_{0}=0.5$, $l=2$
and $\protect\mu =0.1$ ($P\approx 0.5734$). (a1,a2): The production of a
symmetric GS mode by changing $W_{0}$ from }${\protect\small 0.5}$%
{\protect\small \ to $0.1$ and the corresponding power curve $P(z)$.
(b1,b2): The production of a symmetric GS mode by changing $\protect\lambda $
from 0.3 to $0.1$ and the corresponding power curve $P(z)$. The following
panels display the production of asymmetric GS modes from the initial state
corresponding to $W_{0}=0.5$, $l=2$ and $\protect\mu =0.3$ ($P\approx 0.5232$%
). (c1,c2): The production of an asymmetric GS mode by changing $W_{0}$
from }${\protect\small 0.5}${\protect\small \ to $0.1$. (c3,c4): The
corresponding power curve $P(z)$ and asymmetry characteristic $\protect%
\theta (z)$. (d1,d2): The production of an asymmetric GS mode by varying $%
\protect\lambda $ from }${\protect\small 0.2235}${\protect\small \ to $0$.
(d3,d4): The corresponding power curve $P(z)$ and asymmetry characteristic $%
\protect\theta (z)$. }}
\label{jifa}
\end{figure}

We here consider the imaginary potential $\widehat W_2(x,z)$ given by Eq.~(\ref{W2g}).
 First, for symmetric GS modes, keeping $\widehat \lambda $ fixed and replacing $%
\widehat W_{0}(z)$ with $\chi(z)$ given by Eq.~(\ref{lambda}), we can
adiabatically transform the initial state from $\chi_{1}=0.5$ to $%
\chi_{2}=0.1$, as shown in Fig.~\ref{jifa}(a1).
Figure~\ref{jifa}(a2) displays the corresponding power curves $P(z)$.
defined as per Eq.~(\ref{P}). It is seen that an initially stable symmetric
GS mode with a higher power is transformed into another symmetric one with a
lower power. Besides, for the same initial state, if we consider the varying coupling
$\widehat\lambda (z)$, from $\chi_{1}=0.3$ to $%
\chi_{2}=0.1$, a stable symmetric GS mode is carried over into another
symmetric one with the constant power, i.e., $\mathrm{d}P/\mathrm{d}z=0$
(see Eq.~(\ref{dp}) and Figs.~\ref{jifa}(b1,b2)). Therefore, the symmetric
sates can only be converted into another symmetric one, in the framework of
system (\ref{phipsij}).

Next, a question is if an asymmetric state can be converted into a symmetric
one. First, for fixed $\widehat\lambda $, an initially stable asymmetric state is
transformed into an asymmetric one with residual oscillations, if $\widehat W_{0}(z)$
is taken as per Eq.~(\ref{lambda}) with $\chi_{1}=0.5$ to $\chi_{2}=0.1$ (see Figs.~\ref{jifa}(c1,c2)). The corresponding power curve $P(z)$
and asymmetry characteristic $\theta (z)$ are displayed in Fig.~\ref{jifa}%
(c3,c4). Powers $P_{1}$ and $P_{2}$ feature oscillations, while the total
power $P$ remains constant. If we only consider the varying inter-core coupling  $\widehat\lambda (z)$
with $\chi_{1}=0.2235$ to $\chi_{2}=0$, the stable asymmetric state is transformed into a stable symmetric
one, as shown in Figs.~\ref{jifa}(d1-d4). A condition necessary for the
implementation of this conversion is $P_{1}(z_{\mathrm{max}}/10)=P_{2}(z_{%
\mathrm{max}}/10)$. It is observed in Fig.~\ref{jifa}(d4) that, at $z\geq z_{%
\mathrm{max}}/10=10$, a stable symmetric state appears with $\theta =0$.

Similar results can be obtained for the imaginary potential $\widehat W_{1}(x,z)$ given by Eq.~(\ref{W1g}),
which are not presented here in detail.

\section{Physical interpretations}
In Fig.~\ref{pt}, we show the profiles of real and imaginary parts of the $\PT$-symmetryic potentials and the $\PT$ symmetry breaking of the linear spectral problem (\ref{linears}) with differential
gain-and-loss distributions given by Eqs.~(\ref{W1}) and (\ref{W2}). In Fig.~\ref{GS_ssb}, we exhibit the boundary for the breaking of the inter-core symmetry $P_{\mathrm{cr}}(\protect\lambda )$ and $\protect\mu _{\mathrm{cr}}$. Total power $P$ versus propagation constant for GS modes is shown in Fig.~\ref{mup12}. The important results are displayed in Fig.~\ref{GS_w0p}, i.e., the stability and boundary of the breaking of the
inter-core symmetry for GS modes in $(W_{0},P)$-space. It is found that the instability of $\PT$-symmetric solutions takes place prior to the inter-core symmetry breaking.
And, stable inter-core-symmetric GS solutions may remain stable while the $\PT$ symmetry is broken.
Fig.~\ref{Ptheta} shows the asymmetry characteristic $\protect\theta $ (see Eq.~(\protect\ref{asym})) in $(W_{0}, P)$-space. And Fig.~\ref{ssbj} exhibits the stability and boundary of the breaking of the inter-core symmetry for DM modes in $(W_{0},P)$-space. Besides, Figs.~\ref{GS_evo}, \ref{jsym} and \ref{jasym} show the profiles of GS and DM modes as well as the corresponding numerically simulated evolution.
Fig.~\ref{v0} displays the total power versus propagation constant for GS modes in the case of the pure imaginary localized potential $W_2(x)$.
Finally, the transformation of GS modes in the framework of Eq. (\protect\ref{lambda}) is shown in Fig.~\ref{jifa}.

\section{Conclusions and discussions}

In this work, we have introduced the one-dimensional system of linearly
coupled NLS equations with the cubic self-attraction, harmonic-oscillator
trapping potential, and two different types of the spatially odd dissipative
potential (gain-loss distribution), localized and delocalized.
The system
models the light transmission in a dual-core planar waveguide (coupler) with
the Kerr nonlinearity and effective $\mathcal{PT}$-symmetric potential.
By means of numerical methods, symmetric and asymmetric GS (ground-state) and
DM (dipole-mode) solitons have been found.
Due to the action of the harmonic-oscillator trap, it is possible to find stable dipole modes in dissipative
systems. The asymmetric states are generated
from symmetric ones by the SBB (ground-state) of the supercritical type.
The novelty of results in the system is interplay between breakings of
the $\PT$ and inter-core symmetries.
In order to exhibit the feature, the critical value of the power, $P_{\mathrm{cr}}$, at which the breaking of the
inter-core symmetry takes place, is found as a function of strength $W_{0}$
of the gain-loss distribution.
The stability of symmetric and asymmetric GS
and DM modes, affected by the SBB and $\mathcal{PT}$-symmetry breaking, is
investigated by dint of the linear-stability analysis and direct
simulations.
Different from the conservative counterpart of the system, with
$W_{0}=0$, the instability of the symmetric GS solutions commences prior to
the onset of the inter-core symmetry breaking, due to the effect of the
gain-loss distribution. Near the $\mathcal{PT}$-symmetry breaking point,
instability of the symmetric solution sets in at smaller values of the
power. Surprisingly, stable inter-core-symmetric GS solutions may remain stable while the $\PT$ symmetry is broken.
In contrast to the conservative
system, in which all asymmetric states created by the supercritical SBB are
stable, in the present system, which includes the imaginary potential,
asymmetric GS and DM are only partly stable.
We have also investigated the
SBB of symmetric and asymmetric GS modes in the system including a localized
pure imaginary potential, in the absence of the real potential. In this
case, the SBB is of the subcritical type.
And stable solitons can still be found in dissipative systems.
Finally, the transformation of
symmetric and asymmetric GS modes has been studied, following variation of
the strength of the imaginary potential $W_{0}$ or inter-core coupling
constant $\lambda $ along the propagation distance, which demonstrate that the GS modes can be transformed from the asymmetric shape
into the symmetric one, under appropriate conditions.
These results may  provide the theoretical support for the optical experiments in a dual-core planar waveguide with $\PT$-symmetric
potential.

This work suggests an interesting extension for the study of SBB of
fundamental and vortical two-component solutions in the effectively
two-dimensional spatiotemporal dual-core system \cite{Dror} with a $\mathcal{%
PT}$-symmetric potential.
\begin{equation}
\begin{split}
i\phi _{z}& =-\frac{1}{2}\nabla^2\phi+F(|\phi |^{2}, |\psi |^{2})\phi + [V({\bf r})+iW({\bf r})]\phi -\kappa \psi, \\
i\psi _{z}& =-\frac{1}{2}\nabla^2\psi+F(|\phi |^{2}, |\psi |^{2})\psi +[V({\bf r})+iW({\bf r})]\psi -\kappa \phi.
\end{split}
\label{phipsi-02}
\end{equation}
It may also be relevant to consider a
generalization of the system including an optical-lattice potential, which
may give rise to other asymmetric soliton modes.



\vspace{0.25in}
{\small\noindent

\noindent {\bf Authors' contributions.} J.S.: Conceptualization, Methodology, Investigation, Analysis, Writing-original draft.
B.A.M. and Z.Y.: Conceptualization, Methodology, Formal Analysis, Supervision, Funding acquisition, Writing-reviewing and editing.

\noindent {\bf Conflict of interest declaration.} We declare we have no competing interests.

\noindent {\bf Funding.} Z.Y. acknowledges support from the National Nature Science Foundation of China under Grant No. 11925108.
The work of B.A.M. is supported, in part, by the Israel Science Foundation through
grant No. 1695/22. }

\end{document}